\documentclass[12pt]{article}
\usepackage{amsmath,amssymb}
\usepackage{eucal}
\usepackage[dviwindo]{graphicx}
\usepackage[dviwindo]{graphics}

\numberwithin{equation}{section}
\newcommand{\ben}{\begin{eqnarray}}
\newcommand{\een}{\end{eqnarray}}
\newcommand{\la}{\label}
\begin{document}

\title{Exact Solutions of Regge-Wheeler Equation}

\vskip 1.5truecm

\author{P.~P.~Fiziev\thanks{Department of Theoretical Physics, University of Sofia,
E-mail:\,\,\,fiziev@phys.uni-sofia.bg}}

\date{}
\maketitle

\begin{abstract}

The Regge-Wheeler equation describes the axial perturbations of
Schwarz- schild metric in linear approximation. We present its exact
solutions in terms of the confluent Heun's functions, the basic
properties of the general solution, novel analytical approach and
numerical techniques for study of different boundary problems which
correspond to quasi-normal modes of black holes and other simple
models of compact objects. We depict in more detail the exact
solutions of Regge-Wheeler equation in the Schwarzschild black hole
interior and on Kruscal-Szekeres manifold.

\end{abstract}

%
%

\section{Introduction}

The well known Regge-Wheeler equation (RWE)
\ben  \partial_t^2 \Phi_{s,l} +\left(-\partial^2_{x}
+V_{s,l}\right)\Phi_{s,l} = 0 \la{RW}\een
describes the axial perturbations of Schwarzschild metric in linear
approximation and plays an important role in modern perturbation
treatment of Schwarzschild black hole (SBH) physics. Its study has a
long history and significant achievements \cite{QNM}.

The effective potential in RWE (\ref{RW}) reads
$$V_{s,l}(r)=\left(1-{{1}\over r}\right)\left({{l(l+1)}\over {r^2}}
+{{1-s^2}\over {r^3}} \right).$$
The area radius $r\geq 0$ can be expressed explicitly as a function
of the "tortoise" coordinate
$x\!=\!r\!+\!r_{\!{}_{Sch}}\ln\left(|r/r_{\!{}_{Sch}}\!-\!1|\right)$
using Lambert W-function: $r=W(\pm e^{x-1})+1$. In the last formula
sign $"+"$ stands for SBH exterior $r\in (1,\infty)$ and sign $"-"$
-- for the SBH interior $r\in (0,1)$. Hereafter we are using units
in which the Schwarzschild radius $r_{\!{}_{Sch}}\!=\!2M\!=\!1$. The
most important case $s\!=\!2$ describes the gravitational waves and
is our main subject here.

The ansatz $\Phi_{s,l}(t,r)=R_{\omega,s,l}(r) e^{i\omega t}$ brings
us to the stationary problem in the outer domain $r>1$:
\ben
\partial^2_x R_{\omega,s,l}+\left(\omega^2-V_{s,l}\right)
R_{\omega,s,l}=0. \la{R}\een
Its {\em exact} solutions were described recently \cite{F} in terms
of the confluent Heun's functions $HeunC\!\left(\alpha, \beta,
\gamma, \delta,\eta,r \right)$ \cite{Heun}.

In the exterior domain the variable $r$ plays the role of 3D-space
coordinate and the variable $t$ -- the role of exterior time of a
distant observer. Because of the change of the signs of the
components $g_{tt}=-1/g_{rr}=1-1/r$ of the metric, in the inner
domain the former Schwarzschild-time variable $t$ plays the role of
radial space variable $r_{in}=t\in (0,\infty)$ and the area radius
$r\in(1,0)$, plays the role of time variable. As a result, the
Regge-Wheeler "tortoise" coordinate $x\in (-\infty,0)$ presents a
specific time coordinate in the inner domain. For the study of the
solutions of \eqref{RW} in this domain it is useful to stretch the
last interval to the standard one by a further change of the
time-variable: $x\to t_{in}=x-1/x\in (-\infty,\infty)$.

These notes are important for the physical interpretation of the
mathematical results. In particular, the natural form of the
interior solutions of \eqref{RW} is
$$\Phi_{\omega,s,l}^{in}(r_{in},t_{in})=e^{i\omega r_{in}}
R_{\omega,s,l}\left(r(t_{in})\right),$$ where
$$r(t_{in})\!=\!W\!\left(-exp\left( {{t_{in}}\over{2}}\!-\!1\!-\!
\sqrt{\left({{t_{in}}\over{2}}\right)^2\!+\!1}\,\right)
\right)\!\!+\!1.$$
The dependence of this solution on the interior radial variable
$r_{in}$ is simple. Its dependence on the interior time $t_{in}$ is
governed by the \eqref{R} with interior-time-dependent potential
$V_{s,l}$. In despite of this unusual feature of the solutions in
the SBH interior, this way we obtain a basis of functions, which are
suitable for the study of the corresponding linear perturbations.

\section{Exact Solutions of Regge-Wheeler Equation in Terms of Confluent Heun's functions}

In $r$ variable the RWE reads:
\ben \la{P} {{d^2R_{\omega,s,l}}\over{dr^2}}+
{1\over{r(r-1)}}{{dR_{\omega,s,l}}\over{dr}}+\\
{{\omega^2r^4-l(l+1)r^2+\left(l(l+1)+s^2-1\right)r+1-s^2}\over
{r^2(r-1)^2}}R_{\omega,s,l}=0.\nonumber
 \een
The anzatz $R_{\omega,s,l}(r)=r^{s+1}(r-1)^{i\omega} e^{i\omega
r}H(r)$ reduces it to the confluent Heun equation \cite{Heun}:
\ben {{d^2H}\over{dr^2}}+\left(\alpha+{{\beta+1}\over{r}}+
{{\gamma+1}\over{r-1}}\right){{dH}\over{dr}}+\hskip 5.3truecm\nonumber\\
{1\over{r(r-1)}} \left( \Big(\delta+{1\over
2}\alpha(\beta+\gamma+2)\Big) r+\eta+{\beta\over 2}+{1\over
2}(\gamma-\alpha)(\beta+1) \right)H=0\la{H}\een
with the specific parameters: $\alpha=2i\omega, \beta=2s,
\gamma=2i\omega, \delta=2\omega^2, \eta=s^2-l(l+1)$ and three
singular points in the whole complex plane $\mathbb{C}_r$ \cite{F}.
Two of them: $r=0$ and $r=1$ are regular and can be treated on equal
foods. The third one: $r=\infty$ is an irregular singular point.
Note that after all the horizon $r=1$ turns to be a singular point
in the above sense, nevertheless  with respect to the algebraic
invariants of the Riemann's curvature tensor it does not define a
singular surface in the very Schwarzschild space-time manifold
$\mathbb{M}^{(1,3)}\{g_{ij}(t,r,\theta,\phi)\}$.

Correspondingly, one can write down three pairs of exact local
solutions -- one regular and one singular -- around each singular
point $X=0,1,\infty$:

1) Frobenius type of solutions around the points $r=0$ and $r=1$:
\begin{subequations}\label{R0:abc}
\ben \Phi_{\,\,\omega,s,l}^{(0)\pm}(t,r)= e^{i \omega
t}R_{\,\,\omega,s,l}^{(0)\pm}(r),
\hskip 8.9truecm \la{R0:a}\\
R_{\,\,\omega,s,l}^{(0)\pm}(r)= r^{s+1} e^{ i\omega r
+i\omega\ln(1-r)} \begin{cases}
H_{\,\,\omega,s,l}^{(0)+}(r),\,\\
HeunC\!\left(2i\omega, 2s, 2i\omega, 2\omega^2,s^2-l(l+1),r \right)
;\end{cases}  \hskip .6truecm\la{R0:b}\een
\end{subequations}
\begin{subequations}\label{R1:ab}
\ben \Phi_{\,\,\omega,s,l}^{(1)\pm}(t,r)=
e^{i \omega t}R_{\,\,\omega,s,l}^{(1)\pm}(r),\hskip 8.7truecm \la{R1:a}\\
R_{\,\,\omega,s,l}^{(1)\pm}(r)\!=\!r^{s+1} e^{i\omega r \pm
i\omega\ln(r-1)}\times\hskip 7.9truecm \nonumber\\\hskip 3.truecm
HeunC\left(-2i\omega,\pm 2i\omega,2s,
-2\omega^2,2\omega^2+s^2-l(l+1),1-r \right)\!;\la{R1:b}\een
\end{subequations}

\begin{figure}[htbp] \vspace{5.5truecm}
\includegraphics{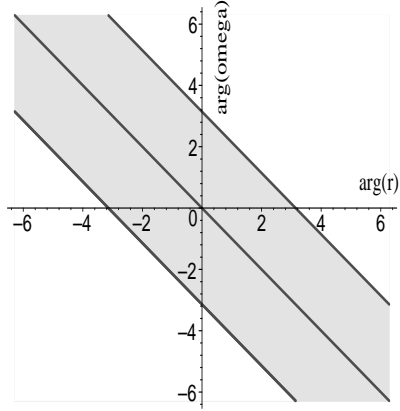}
\vskip .5truecm
\caption{Domains of different behavior of
the functions $R_{\,\,\omega,s,l}^{(\infty)\pm}(r)$.
    \hskip -1.truecm}
    \label{Fig1}
\end{figure}

2) Asymptotic Tome type of solutions around the point $r=\infty$:
\begin{subequations}\label{Rinf:ab}
\ben \Phi_{\,\,\omega,s,l}^{(\infty)\pm}(t,r)=
e^{i \omega t}R_{\,\,\omega,s,l}^{(\infty)\pm}(r),\hskip 8.5truecm \la{Rinf:a}\\
   R_{\,\,\omega,s,l}^{(\infty)\pm}(r)\sim{{e^{\mp i\omega r -i\omega\ln r}
}} \sum_{\nu\geq 0}{{a_\nu^{\pm}}\over{r^\nu}},\,\,\,\,a_0^\pm=1.
\hskip 6.3truecm\la{Rinf:b}\een
\end{subequations}
These solutions describe ingoing and outgoing waves, associated with
the corresponding singular point, according to the following scheme.
\vskip 1.truecm
\begin{table}[here]~\vspace{-1.truecm}
\caption{\label{T1}In-out properties of local solutions.}
\begin{center}
\begin{tabular}{cllllrrrr}
Ingoing solutions:&\,\,\,&$\hskip 2.truecm
$&Outgoing solutions:&\,\,\,&\\
\\
$\Phi_{\,\,\omega,s,l}^{(0)+}(t,r):$&$0\leftarrow$&$
$&$\Phi_{\,\,\omega,s,l}^{(0)-}(t,r):$&$0\rightarrow$&\\
$\Phi_{\,\,\omega,s,l}^{(1)+}(t,r):$&$1\leftarrow$&$
$&$\Phi_{\,\,\omega,s,l}^{(1)-}(t,r):$&$1\rightarrow$&\\
$\Phi_{\,\,\omega,s,l}^{(\infty)+}(t,r):$&$\rightarrow\infty$&$
$&$\Phi_{\,\,\omega,s,l}^{(\infty)-}(t,r):$&$\leftarrow\infty$&\\
\end{tabular}
\end{center}
\la{table1}
\end{table}
\vskip .4truecm
\noindent The solutions own the following properties:
\begin{subequations}\label{r_lim:ab}
\ben \lim_{r\to X}
|\Phi_{\,\,\omega,s,l}^{(X)+}(t,r)|&=&\infty,\label{r_lim:a}\\
\lim_{r\to X} |\Phi_{\,\,\omega,s,l}^{(X)-}(t,r)|&=&0,
\label{r_lim:b}\een
\end{subequations}
with the important justification in the case $|\omega|>0$, as
described by the equations
\begin{subequations}\label{limRinf:ab}
\ben \lim_{|r|\to \infty}
|R_{\,\,\omega,s,l}^{(\infty)+}(r)|=\begin{cases} \infty,\,\,\,
&\hbox{if}\,\,\, \arg(r)+\arg(\omega)\in(0,\pi)/
\hskip -.4truecm\mod 2\pi;\label{limRinf:a}\\
0, \,\,\, &\hbox{if}\,\,\, \arg(r)+\arg(\omega)\in(-\pi,0)/\hskip
-.4truecm\mod 2\pi.
\end{cases}\\
\lim_{|r|\to \infty}
|R_{\,\,\omega,s,l}^{(\infty)-}(r)|=\begin{cases} 0,\,\,\,
&\hbox{if}\,\,\, \arg(r)+\arg(\omega)\in(0,\pi)/
\hskip -.4truecm\mod 2\pi;\label{limRinf:b}\\
\infty, \,\,\, &\hbox{if}\,\,\,
\arg(r)+\arg(\omega)\in(-\pi,0)/\hskip -.4truecm\mod 2\pi,
\end{cases}
\een \end{subequations}
and illustrated in Fig.\ref{Fig1}.
The general solution of the RWE equation (\ref{R}) can be
represented in the form
$R_{\omega,s,l}(r)=
C_{X+}R_{\,\,\omega,s,l}^{(X)+}(r)+
C_{X-}R_{\,\,\omega,s,l}^{(X)-}(r)$ with proper constants
$C_{X\pm}$.
The different local solutions are related according the formula
$$R_{\,\,\omega,s,l}^{(X)\pm}(r)= \Gamma_{\,Y+}^{X\pm}(\omega,s,l)
R_{\,\,\omega,s,l}^{(Y)+}(r)+ \Gamma_{\,Y-}^{X\pm}(\omega,s,l)
R_{\,\,\omega,s,l}^{(Y)-}(r).$$
The main obstacle for pure analytical treatment of different
interesting problems, related to the Regge-Wheeler equation, is that
the transition coefficients $\Gamma_{\,Y\pm}^{X\pm}(\omega,s,l)$ for
$X,Y=0,1,\infty$; $X\neq Y$ are not known explicitly.

\section{Different Boundary Problems for the Stationary Regge-Wheeler Equation }
\begin{figure}[htbp] \vspace{5.5truecm}
\includegraphics{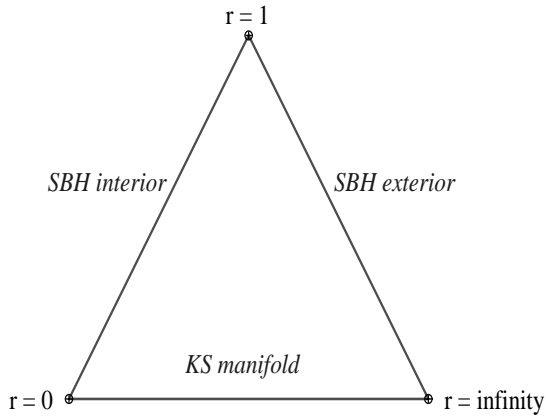}
\vskip .5truecm
\caption{Different types of
two-singular-ends boundary problems.
    \hskip -1.truecm}
    \label{Fig2}
\end{figure}

Due to the existence of three singular points of RWE, we have three
different two-singular-ends boundary problems \cite{Heun} in the
complex plane $\mathbb{C}_r$, Fig.\ref{Fig2}. These describe
different physical problems, related with SBH exterior, SBH interior
and Kruscal-Szekeres (KS) extension of the Schwarzschild solution
\cite{GR}.

Besides we can consider regular-singular-ends boundary problems on
the interval $(r_*,\infty)$, $r_*>1$ which can describe the
perturbation of the metric around compact spherically symmetric
matter objects in rest.

\subsection{Perturbations of the SBH Exterior}

Most well studied are the linear perturbations of the SBH exterior
\cite{RW, QNM}. Since the quasi-normal modes (QNM) are defined as a
stable solutions of RWE (\ref{RW}) which describe pure ingoing waves
both into the horizon $r=1$ and into the space infinity $r=\infty$,
we can describe them using the solutions
$R_{\,\,\omega,s,l}^{(1)+}(r)$ (\ref{R1:b}) of \eqref{R}, which obey
the non-explicit equation $\Gamma_{\infty
-}^{\,1\,\,+}(\omega,s,l)=0$. It yields the QNM frequencies
$\omega=\omega(n,s,l),\,\,\,n=0,1,2,\dots$\,. Our exact approach
gives the spectral condition \cite{F}:
\ben \lim_{|r|\to +\infty}\bigg|\,r^{s+1}\!
HeunC\left(\!-2i\omega,2i\omega,2s,
-2\omega^2,2\omega^2\!\!+\!s^2\!-\!l(l\!+\!1), 1\!-\!|r|
e^{\!-i\left(\!{\pi \over
2}\!+\!\arg(\omega)\right)}\!\right)\!\bigg|\nonumber\\=0.
\la{NumBHspect}\een
It expresses in an {\em explicit} way the condition that the ingoing
at the horizon waves do not contain coming from the infinity waves
and can be solved numerically using the computer package Maple 10.

The equation \eqref{NumBHspect} reproduces with a great precision
the known numerical results, obtained by different methods
\cite{QNM}. For example, for the basic QNM-eigenvalue our
calculations give $\omega(0,2,2)=0.747343368836 + i\, 0.177924631360
$ \cite{F}, thus justifying the previously published results
\cite{QNM}. The first six eigenvalues are shown in Fig.\ref{Fig7}.

The amplitude of all QNM solutions
$R_{\,\,\omega(n,s,l),s,l}^{(1)+}(r)$ increase infinitely around the
horizon and near the space infinity, according to the
\eqref{r_lim:a}. This shows that Regge-Wheeler perturbation theory
is not applicable in the corresponding vicinities of these points.

\subsection{Perturbations of the SBH Interior}
The perturbations of SBH interior were considered qualitatively for
the first time in \cite{Matzner}. We present here an explicit exact
treatment of the problem.

In the classical theory of BH it seems reasonable to exclude from
the most of the considerations the inner domain, because it does not
influence {\em directly} the observable outer domain. Nevertheless,
the excitations of degrees of freedom in the inner domain may be
essential for the consideration of the total BH entropy and for
construction of the quantum theory of BH. The interior excitations
may yield observable effects in the exterior of the BH, due to the
quantum effects \cite{BMR}.

Another reason to study the BH interior was discovered in the recent
article \cite{LBLR}: It was shown that without excision of the inner
domain around the singularity one can improve dramatically the
long-term stability of the numerical calculations. Although the
region of the space-time that is causally disconnected can be
ignored for signals and perturbations traveling at physical speeds,
numerical signals, such as gauge waves or constraint violations, may
travel at velocities larger then that of light and thus leave the
physically disconnected region.

The perturbative treatment is related with the metric's dynamics and
corresponding initial value problem: using the field equation one
has to determine the evolution of the initial data on a proper
Cauchy surface. In the exterior domain solutions of \eqref{RW},
which grow up infinitely with respect to the future direction of the
exterior time $t$ do exist. They are unstable and physically
unacceptable. Therefore such solutions are excluded from any
physical consideration \cite{QNM, RW}.

Then one has to formulate a proper boundary problem for \eqref{R},
which has a reach variety of solutions and may describe different
physical problems. For a correct formulation of a given physical
problem one restricts the class of the solutions using proper
boundary conditions. Since the physics at the boundary influences
strongly the solutions of the problem, the boundary conditions are a
necessary ingredient of the theory and define the physical problem
under consideration. Only after that the physical problem is fixed
and one can prove the stability of the remaining solutions (QNM) and
small perturbations of general form with respect to the
exterior-time evolution \cite{NM}. Our point is to consider
different type of mathematical boundary problems, related with
\eqref{R}.

By analogy with the consideration in the outer domain of SBH, in the
inner one we are considering only solutions, which are stable with
respect to the future direction of the interior time. Therefore we
study the solutions of \eqref{R}, which are regular at the point
$r\!=\!0$. These have a nontrivial spectrum with novel properties.
Since we are working in the framework of perturbation theory, the
presence of "bad" solutions $R_{\,\,\omega,s,l}^{(0)+}(r)$, see
\eqref{R0:b}, which diverge and even are not single-valued, shows
that, in general, two different types of situation can take place:
either the considered here perturbation theory does not work in some
domain around the interior-future infinity, or the full nonlinear
problem has "bad" solutions, which have to be excluded from the
physical considerations using proper additional conditions. It is
impossible to resolve this problem in the framework of the linear
perturbation theory. Nevertheless, it is clear, that the opposite
case of the regular solutions $R_{\,\,\omega,s,l}^{(0)-}(r)$, see
\eqref{R0:b}, has always a well defined physical meaning around
future infinity of the interior time and deserves a corresponding
study. For this purpose we are using a second pair of solutions
around the horizon $r=1$:
\ben \bar R_{\,\,\omega,s,l}^{(1)\pm}(r)\!=\!r^{s\!+\!1} e^{\!\pm
i\omega \left(\ln(1\!-\!r)\!-r\right)} HeunC\!\left(\pm2i\omega,
\pm2i\omega, 2s,
-2\omega^2,2\omega^2\!\!+\!s^2\!\!-\!l(l\!+\!1),1\!-\!r
\right)\!,\hskip 0.4truecm \la{r1}\een
which can be derived from the solutions (\ref{R1:b}), applying
proper transformations of the Heun's functions \cite{Heun}. For
$\omega\neq 0$ the functions $\bar R_{\,\,\omega,s,l}^{(1)\pm}(r)$
(\ref{r1}) define a pair of linearly independent solutions with the
symmetry property
$\bar R_{\,\,\omega,s,l}^{(1)\pm}(r)= \bar
R_{\,\,-\omega,s,l}^{(1)\mp}(r)$.

\begin{figure}[htbp] \vspace{5.5truecm}
\includegraphics{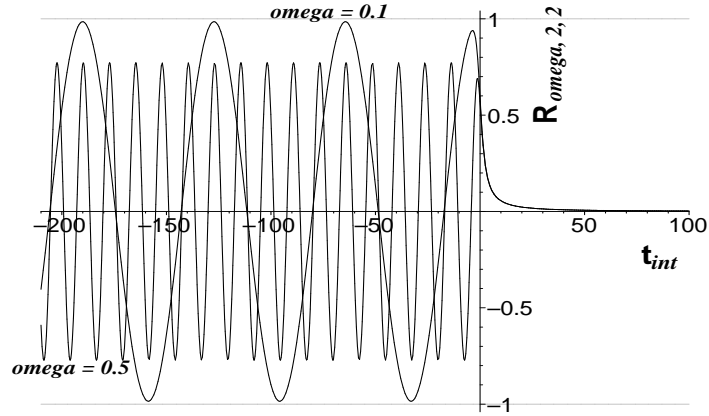}
\vskip .5truecm \caption{Two solutions of the continuous spectrum.
    \hskip -1.truecm}
    \label{Fig3}
\end{figure}

1. For real $\omega^2>0$ the problem has a real continuous spectrum.
The corresponding solutions $R_{\omega, 2, 2}(t_{in})$ are
illustrated in Fig.\ref{Fig3}.  They define a basis of stable {\em
normal modes} for perturbations of the metric in the SBH interior
and can be used for Fourier expansion of perturbations of more
general form.

2. The zero eigenvalue $\omega=0$ yields a degenerate case in which
the two functions (\ref{r1}) coincide. Then we obtain the infinite
series of {\em polynomial} solutions:
\ben
R_{\,\,\omega=0,s,l}^{(0)-}(r)\!=\!{{(l\!-\!s)!}\over{Poch(1\!+\!2s,l\!-\!s)}}
r^{s\!+\!1} JacobiP(l\!-\!s,2s,0,1\!-\!2r)=r^3\!+\!\dots,
\la{e0}\een
which are finite on the whole interval $r\in[0,1]$.  These functions
form an orthogonal basis of stable normal modes. Here
$JacobiP(n,a,b,z)$ stands for the standard Jacobi polynomials and
$Poch(c,d)$ -- for the Pochhammer symbol.

3. The discrete spectrum of the problem at hand corresponds to pure
imaginary values of the parameter $\omega=i\omega_I$ \cite{Matzner,
F}, i.e. to Laplace transform of the perturbations of general type.
Thus, studying the the case $\omega^2<0$ we are constructing a basis
for Laplace expansion of the perturbations of more general form in
the SBH interior.

Then the functions $R_{\,\,\omega,s,l}^{(0)-}(r)$, $\bar
R_{\,\,\omega,s,l}^{(1)\pm}(r)$ and all their parameters are real.
In particular,
\ben\la{R01} R_{\,\,\omega,s,l}^{(0)-}(r)\!=\!
\Gamma_{1+}^{0-}(\omega,s,l) \bar R_{\,\,\omega,s,l}^{(1)+}(r)+
\Gamma_{1-}^{0-}(\omega,s,l) \bar R_{\,\,\omega,s,l}^{(1)-}(r)
\\ \nonumber
\hskip 1.65truecm\sim \cos\alpha\,\bar R_{\,\,\omega,s,l}^{(1)+}(r)+
\sin\alpha\,\bar R_{\,\,\omega,s,l}^{(1)-}(r) \hskip 2truecm
\een
with real transition coefficients $\Gamma_{1-}^{0\pm}(\omega,s,l)$
and real mixing angle $\alpha$, defined via the relation
$\tan\alpha=\Gamma_{1-}^{0-}(\omega,s,l)/
\Gamma_{1+}^{0-}(\omega,s,l)\in \mathbb{R}$ for $\omega=i\omega_I$.

The physical meaning of the mixing angle $\alpha$ is obvious: since
the solutions $\bar R_{\,\,\omega,s,l}^{(1)+}(r)$ describe the
ingoing into the horizon waves and the solutions $\bar
R_{\,\,\omega,s,l}^{(1)-}(r)$ -- the outgoing from the horizon
waves, this angle describes the ratio of the amplitudes of these
waves in their mixture (\ref{R01}). The value $\alpha=0$ corresponds
to absence of outgoing waves (i.e. to the authentic SBH boundary
conditions at the horizon). The value $\alpha= \pi/2$ describes the
case without ingoing waves. The values $\alpha \in (0,\pi/2]$
describe an extension of the simple SBH problem in which the
interior domain has a more general meaning then a pure SBH interior.
Thus, fixing the value of the mixing angle $\alpha \in [0,\pi/2]$ we
are defining completely a more wide class of problems than the
traditional SBH.
\begin{figure}[h]\vskip 1.5truecm
\begin{center}
\hskip -2truecm
\begin{minipage}{11pc}
\includegraphics[width=6pc, viewport=41 1 200 200]{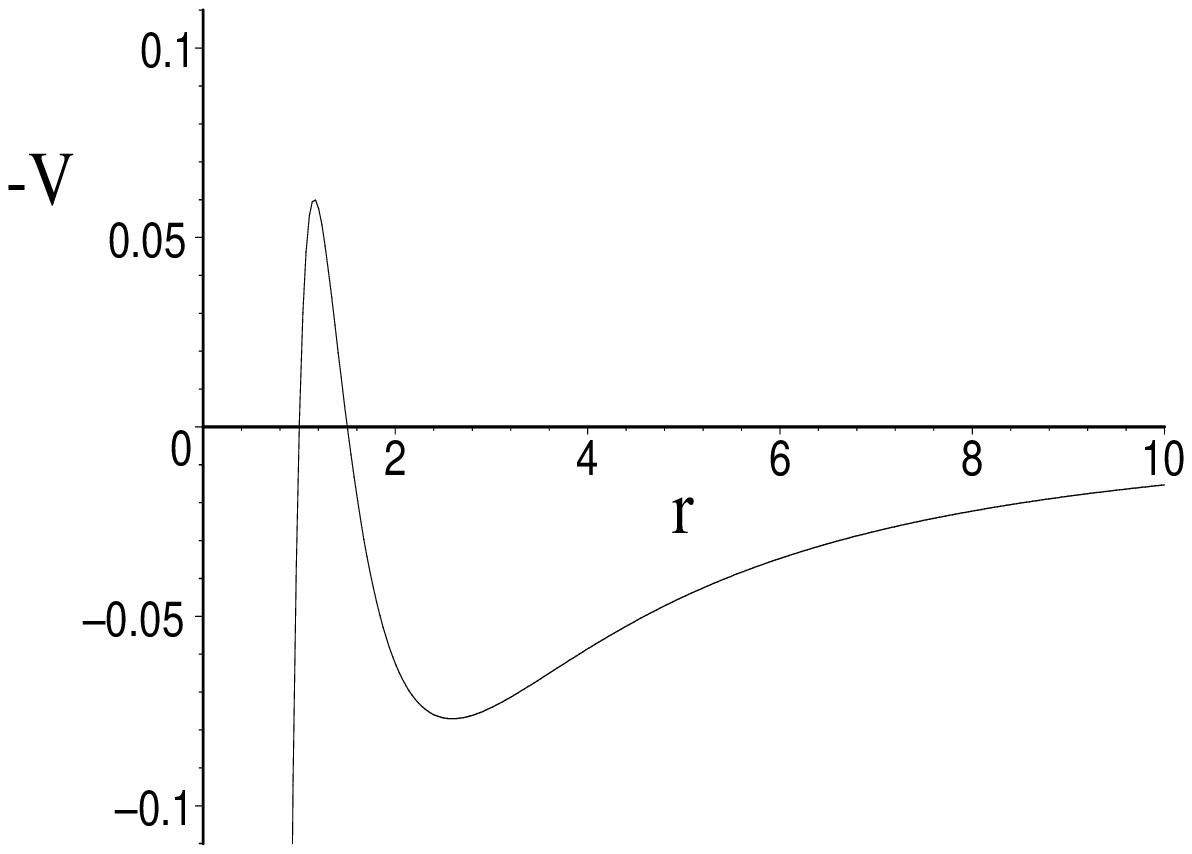}
\caption{\label{Fig4a} The inverse potential $-V(r)$.}
\end{minipage}\hspace{8pc}%
\begin{minipage}{11pc}
\includegraphics[width=6pc, viewport=41 1 190 180]{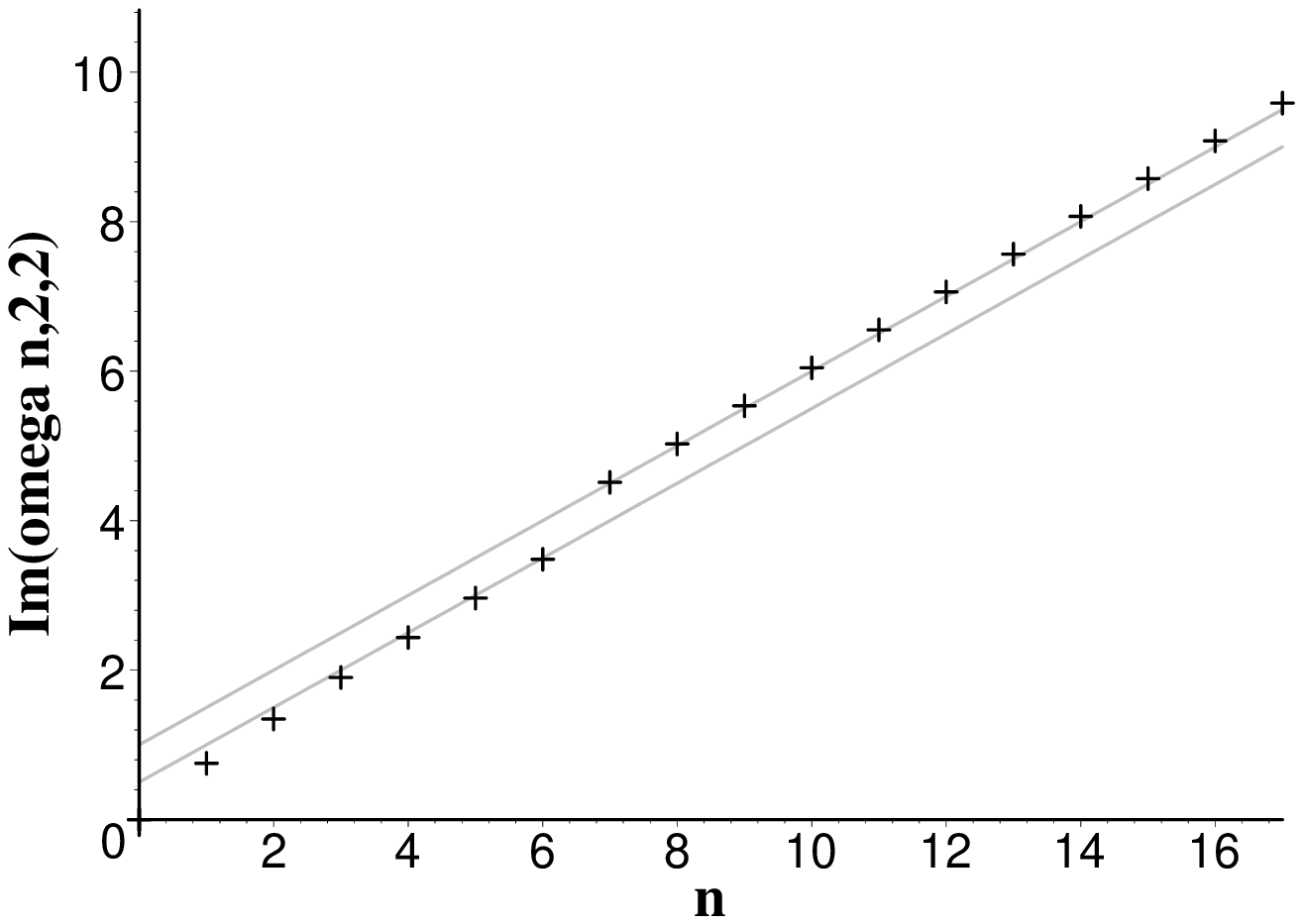}
\caption{\label{Fig4b} The first 18 eigenvalues (for n=0,
2,\dots,17, s=l=2 and $\alpha=0$).}
\end{minipage}
\end{center}
\end{figure}

The discrete imaginary spectrum can be obtained from the spectral
equation:
\ben \left|\begin{array}{cc} R_{\,\,\omega,s,l}^{(0)-}(r_1) &
\cos\alpha\,\bar R_{\,\,\omega,s,l}^{(1)+}(r_1)\!+\! \sin\alpha\,
\bar R_{\,\,\omega,s,l}^{(1)-}(r_1)
\\ R_{\,\,\omega,s,l}^{(0)-}(r_2)
&\cos\alpha\,\bar R_{\,\,\omega,s,l}^{(1)+}(r_2)\!+\! \sin\alpha\,
\bar R_{\,\,\omega,s,l}^{(1)-}(r_2)
\end{array}\right|=0,\hskip .6truecm\la{spectralEq}\een
where $r_{1,2}\in (0,1), r_1\neq r_2$ are two otherwise arbitrary
points. Because of the symmetry property of solutions $\bar
R_{\,\,\omega,s,l}^{(1)\pm}(r)$, it is enough to study only the
solutions $\omega_{n,s,l}$ of the equation (\ref{spectralEq}) with
positive $\omega_I=Im\left(\omega_{n,s,l}\right)>0$.

In the Fig.\ref{Fig4b} we see the first 18 eigenvalues (including
the zero eigenvalue) of the RWE for SBH interior for $s=l=2$ and
$\alpha=0$. The two series: $n=0, 1,\dots6$; and $n=7,\dots$, which
correspond to the two  wells of the  inverse potential $-V(r)$, are
clearly seen. They are placed around the straight lines
$\omega_I=n/2+1/2$ and $\omega_I=n/2+1$, correspondingly.

\begin{figure}[htbp] \vspace{4.5truecm}
\includegraphics{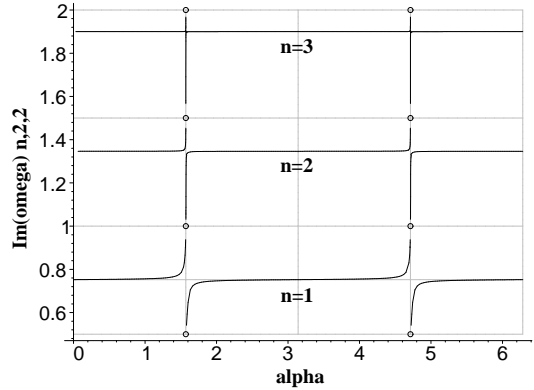} \vskip 1.truecm
\caption{The dependence of $Im\left(\omega_{n,2,2}\right)$ on the
mixing angle $\alpha\in [0,2\pi]$.
    \hskip -1.truecm}
    \label{Fig5}
\end{figure}

In Fig.\ref{Fig5} we see a specific attraction and repulsion of the
eigenvalues $\omega _{n,s,l}(\alpha)$. Such behavior is typical for
the eigenvalue problems related with the Heun's equation
\cite{Heun}. It is well known in the quantum physics. To the best of
our knowledge, this phenomenon is observed for the first time in the
theory of the perturbations of Schwarzschild metric and needs
further study to reach the corresponding physical understanding.

\subsection{Perturbations of the Kruscal-Szekeres Manifold}
To study the stable solutions of the first-order Regge-Wheeler
perturbation theory in this case one has to choose the functions
$$R_{\,\,\omega,s,l}^{(0)-}(r)\!=\! \Gamma_{\infty
+}^{0-}(\omega,s,l) R_{\,\,\omega,s,l}^{(\infty)+}(r)+
\Gamma_{\infty -}^{0-}(\omega,s,l)
R_{\,\,\omega,s,l}^{(\infty)-}(r),$$ which are regular at the
interior-future-infinity $t_{int}=\infty$, and then to impose on
them the condition $\Gamma_{\infty -}^{0-}(\omega,s,l) =0$. It is
needed to annulate the waves, coming from the space infinity
$r=\infty$. Making use of \eqref{R0:abc}, \eqref{Rinf:ab} and
\eqref{limRinf:ab}, it is not hard to obtain the following explicit
form of this condition:
\ben \lim_{|r|\to\infty}\Big|r^{s+1}e^{2i\omega \ln
r}HeunC(2i\omega,2s,2i\omega,2\omega^2,s^2-l(l+1),r)
\Big|_{r=|r|e^{-i({\pi\over 2}+arg\omega)}}\!=\!0.\la{KS}\een
Our numerical study of the case $s=l=2$ shows that this equation has
no finite complex solutions $\omega$ with $|\omega|<\infty$. Hence,
in this case all existing solutions of RWE with no waves coming from
special infinity are growing infinitely when approaching the point
$r=0$, i.e. they are unstable in direction of the interior future.

As an example of such solution one can consider the functions
$R_{\,\,\omega,s,l}^{(1)-}(r)$. According to the \eqref{r_lim:b},
when $Im(\omega)>0$ these functions are finite and continuous in the
whole interval $r\in (0,\infty)$. They tend continuously to zero at
both sides of the horizon $r=1$ (which in this case is a branching
point in $\mathbb{C}_r$) and to infinity -- when approaching the
singular points $r=0$ and $r=\infty$. Unfortunately, these solutions
contain both going to space infinity and coming from there waves.
Our numerical study of the case $s=l=2$ shows that it is impossible
to fulfill the condition
\ben \la{NoCommingWaves}
\Gamma_{\infty
-}^{1-}(\omega,s,l)=\lim_{|r|\to\infty}\left.
{{R_{\,\,\omega,s,l}^{(1)-}(r)}\over{R_{\,\,\omega,s,l}^{(\infty)-}(r)}}
\right|_{r=|r|e^{-i(\pi/2+arg(\omega))}} =\hskip 5.truecm\\
\lim_{|r|\to\infty}r^{1+s}HeunC(-2i\omega,
2i\omega,2s,-2\omega^2,s^2\!-\!l(l\!+\!1)\!+2\omega^2,1-r)
\Big|_{r=|r|e^{-i(\pi/2+arg(\omega))}}\nonumber\\ =0,\nonumber \een
which excludes the coming from the space infinity waves and is
expecting to yield the corresponding spectrum of the frequencies
$\omega$.

Another example are solutions $R_{\,\,\omega,s,l}^{(1)+}(r)$,
considered in the whole interval $r\in (0,\infty)$. Our
consideration in Section 3.1 shows that in this case one is able to
exclude the coming from space infinity waves, obtaining this way the
standard QNM frequencies. These solutions go to infinity both at the
horizon and at the singular points $r=0$ and $r=\infty$. Hence, they
are not stable in direction of the interior future and are not
suitable for constructing perturbation theory in vicinity of all of
the three singular points of the RWE.

\subsection{Perturbations of the Exterior of Compact Matter Objects}
Here we consider a one-parameter family of real intervals $(r_{\!*},
\infty)\subset (1,\infty)$ with Dirichlet's boundary condition at
the left end:
$ \Phi_{\omega,s,l}(t,r_{\!*})=0$.
Physically this means that the waves are experiencing a {\em total}
reflection at the spherical surface with the area radius
$r=r_{\!*}>1$, instead of going freely through it without any
reflection, as in the case of black holes.

\begin{figure}[h]
\vskip .8truecm
\begin{center}
\hskip 2truecm
\begin{minipage}{11pc}
\includegraphics[width=6pc, viewport=-50 -70 70 100]{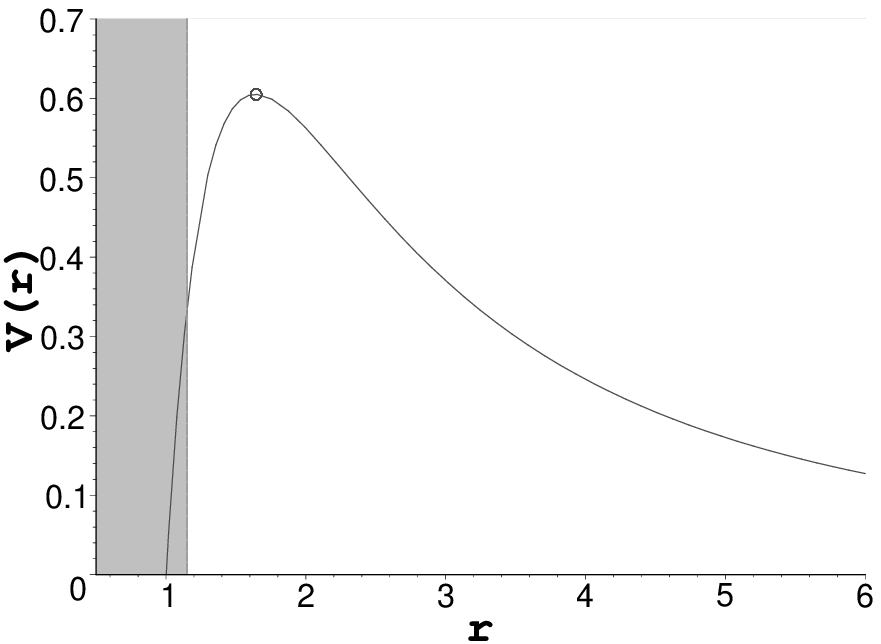}
\caption{\label{Fig6}The potential $V(r)$, truncated at $r_*>1$.}
\vspace{5pc}
\end{minipage}\hspace{3pc}%
\begin{minipage}{16pc}
\includegraphics[width=4.6pc, viewport=-41 -60 90 110]{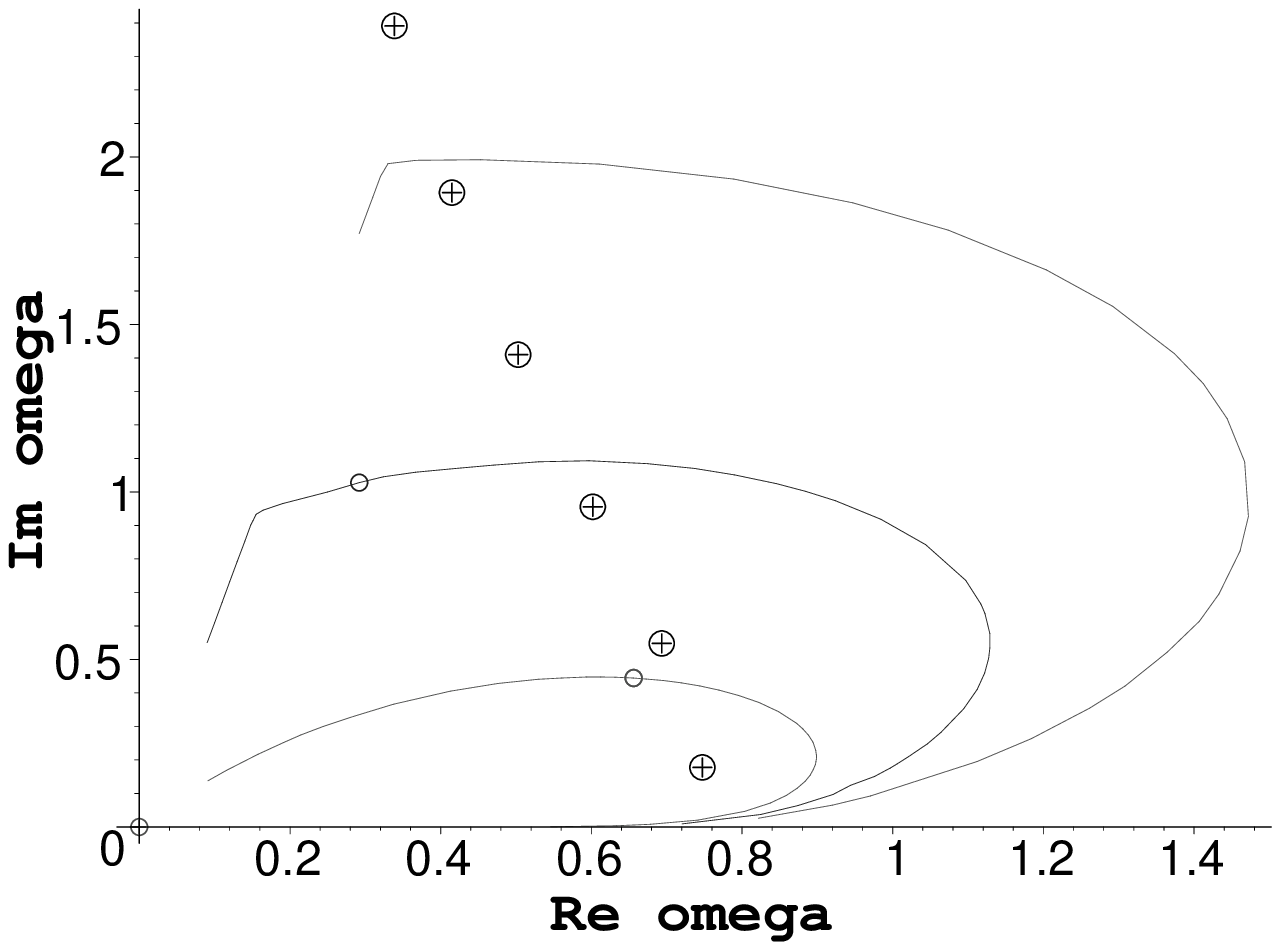}
\caption{\label{Fig7}The dependence of the eigenvalues: $n=0,1,2;
s=l=2$ of a compact body on the area-radius $r_*$ of the totaly
reflection of the perturbations, and the first six QNM of SBH.}
\end{minipage}
\end{center}
\vspace{-3pc}
\end{figure}

One can consider this surface as a boundary of some {\em massive}
spherically symmetric body.  Owing to the Birkhoff theorem, in
general relativity the gravitational field of any massive body with
Keplerian mass $M$ outside the radius $r_{\!*}$ coincides with the
corresponding field of a black hole. Hence, the spreading of the
waves in the outer domain of the massive body is governed by the
Regge-Wheeler equation (\ref{RW}), too, but the boundary conditions
must be different. The corresponding effective potential
$V_{s,l}(r)$ in the Regge-Wheeler equation (\ref{RW}) with
Dirichlet's boundary condition and the corresponding spectrum are
shown in Fig.\ref{Fig6} and Fig.\ref{Fig7}.
The spectral equation, solved numerically to reproduce the
Fig.\ref{Fig7}, expresses the requirement to have at big distances
only going to infinity waves and reads:
\ben \Bigg|\lim_{|r|\to +\infty}\Bigg(
e^{2i*\omega\ln\left(\!|r|e^{-i(\pi/2
+\arg(\omega))}\!-\!1\!\right)}\times\hskip 7.truecm\nonumber\\
{{HeunC\left(\!-2i\omega,+2i\omega,2s,-2\omega^2,2\omega^2\!+\!s^2\!
-l(l\!+\!1),1\!-\!|r|e^{-i(\pi/2+\arg(\omega))}\!\right)} \over
{HeunC\left(\!-2i\omega,-2i\omega,2s,-2\omega^2,2\omega^2\!+\!s^2\!
-\!l(l\!+\!1),1\!-\!|r|e^{-i(\pi/2+\arg(\omega))}\!\right)}
}\Bigg)-\\
e^{2i*\omega\ln\left(\!r_{\!*}\!-\!1\!\right)}\times
{{HeunC\left(\!-2i\omega,+2i\omega,2s,-2\omega^2,2\omega^2\!+\!s^2\!
-l(l\!+\!1),1\!-\!r_{\!*}\!\right)} \over
{HeunC\left(\!-2i\omega,-2i\omega,2s,-2\omega^2,2\omega^2\!+\!s^2\!
-\!l(l\!+\!1),1\!-\!r_{\!*}\!\right)}}\Bigg|=0.\hskip
-.35truecm\nonumber
 \la{NumPPspect}\een

A more realistic model for real spherically symmetric bodies may be
obtained considering as a boundary condition a partial reflection
and partial penetration of the waves through the body's surface.

Such results may help to clarify the physical nature of the observed
very heavy and very compact dark objects in the universe. Our
consideration gives a unique possibility for {\em a direct}
experimental test of the existence of space-time {\em holes}. The
study of spectra of the corresponding waves, propagating around
compact dark objects, may give indisputable evidences, whether there
is space-time hole inside such an invisible object.

\section{Conclusion}

We have demonstrated that the Heun's functions are an adequate
powerful tool for description of the Regge-Wheeler linear
perturbations of the Schwarzschild gravitational field not only
outside the SBH horizon, but in the interior domain as well. They
give, too, the possibility for exact treatment of different boundary
problems for the Regge-Wheeler equation, including ones, which are
related with matter spherically symmetric compact objects. Using
explicit technics we have demonstrated the limitations of the
Regge-Wheeler perturbation theory of SBH. Since this theory has no
complete basis of solutions, which are finite everywhere in the
range $r\in(0,\infty)$, or at least at finite distances, one can
conclude that Regge-Wheeler perturbation theory is not adequate for
an approximate description of the real properties of the full
nonlinear problem, which describes small deviations of the
Schwarzschild metric. Thus we have seen explicitly that this theory
may have only a limited range of applications. The search for a
better type of perturbation theory, or the full nonlinear treatment
of the small perturbations of the SBH remains an open problem.

\vskip .5truecm
\centerline{\bf Acknowledgments}

\vskip .3truecm

The author is thankful to Kostas Kokkotas for useful discussion of
the exact solutions of RWE and SBH interior solutions and to Luciano
Rezzolla for the stimulating discussion of the recently found
numerical techniques for treatment of BH problems without excision
of the interior domain during the XXIV Spanish Relativity Meeting,
E.R.E. 2006.

This article was supported by the Scientific Found of Sofia
University, Contract 70/2006, by its Foundation "Theoretical and
Computational Physics and Astrophysics" and by the Scientific Found
of the Bulgarian Ministry of Sciences and Education, Contract VUF
06/05.

\end{document}